\documentstyle{article}

\begin{document}

\title{Bi--Hamiltonian manifolds, quasi-bi-Hamiltonian
systems and separation variables}

\author{G. Tondo \\
Dipartimento di Scienze Matematiche, Universit\`a degli
Studi di Trieste, \\ Piaz.le
Europa 1, 
I-34127  Trieste, Italy. \\
E-mail: tondo@univ.trieste.it \\
 C. Morosi \thanks{ Supported by  the GNFM of the Italian
CNR.}
\\ Dipartimento di Matematica, Politecnico di Milano, \\
Piazza L. Da Vinci 32, 
I-20133 Milano, Italy. \\
E-mail: carmor@mate.polimi.it }

\date{ }

\maketitle

\begin{abstract}\noindent
We discuss from a
bi-Hamiltonian point of view  the Hamilton--Jacobi
separability of a few
dynamical systems. They are shown to  admit, in  their
natural phase space,
 a quasi--bi--Hamiltonian formulation of Pfaffian type.
This property allows us to
straightforwardly recover a set of separation variables for
the corresponding
Hamilton--Jacobi equation.

\end{abstract}

\section {Introduction} \label{sec:intro}

 The notion of quasi-bi-Hamiltonian (QBH) systems has been
introduced  quite recently \cite{francesi2}; it originates from
the study of dynamical systems which  are associated with two
compatible Poisson tensors $ (P_0, P_1)$, but do not admit   a
bi--Hamiltonian formulation, in the given
phase space, w.r.t. to these tensors. 
\par
 At least for a class of such systems (the so-called
Pfaffian QBH)  it has been shown
\cite{MT} how to   construct  a set of separation variables
for the corresponding  Hamilton-Jacobi
equation, so that a Pfaffian QBH system can be integrated
by quadratures.  In our  opinion, this
relation between QBH formulation and separation of
variables is  remarkable.
\par
On the other hand, since at
present there is not, at the best of our  knowledge, a
satisfactory general scheme encompassing these
properties,  we believe that it may be preliminarily useful
to study some concrete
examples  of such systems; as a matter of fact, it turns
out  that the QBH
formulation is shared by a few  classical separable systems
(as for other cases
previously considered in the  literature, see
\cite{francesi1,Bl1,Bl2,MTcal}).
In the next section we briefly review the main definitions
and   results to
be used in Sect. 3, where the following systems are
presented: the Kepler problem with a
homogeneous force, the Euler problem  with two fixed
centers and an elastic force, the motion on the
ellipsoid in  an elastic potential and a solvable $n$-body
problem  introduced in
\cite{Cal}.

\section{Bi-Hamiltonian manifolds and
quasi--bi--Hamiltonian systems}  \label{sec:BHm}

A bi--Hamiltonian (BH) manifold \cite{Magri1} is a smooth
manifold $M$ endowed with two compatible
Poisson tensors $P_0$, $P_1:T^*M\mapsto TM$ ($TM$
and $ T^*M$ being the tangent and cotangent bundle of $M$,
respectively). As it  is known,
if
$M$ is even dimensional (dim
$M= 2n$) and at least a  Poisson tensor, say $P_0$, is
invertible, then $N=P_1
~P_0^{-1}$ is a  Nijenhuis tensor
\cite{MagriL} (a hereditary operator in the terminology of
\cite{FoFu}).
  In particular, if $N$ is maximal, i.e.,  it has $n$
functionally independent eigenvalues
$(\lambda_1,\ldots ,\lambda_n)$, one can introduce a set of
canonical coordinates
$(\mbox{\boldmath  $\lambda$};\mbox{\boldmath $ \mu$)}$
 ($ \mbox{\boldmath
$\lambda$}:=(\lambda_1,\ldots,\lambda_n);
\mbox{\boldmath  $\mu$}:=(\mu_1,\ldots,\mu_n)$) referred to
as
Darboux-Nijenhuis coordinates  such that
$P_0$, $P_1$ and $N$ take the matrix  form

\begin{equation} \label{eq:PND} P_0= \left[
\begin{array}{cc}
  0& I\\
 - I&0
\end{array} \right] \ , \qquad P_1 = \left[
\begin{array} {cc}
0&\Lambda\\
 -\Lambda &0
\end{array} \right] \ , \qquad N=\left[
\begin{array}{cc}
\Lambda&0 \\
 0&\Lambda
\end{array} \right] \ ,
\end{equation} where $ I$ denotes the $n \times n$ identity
matrix and $
 \Lambda:=diag(\lambda_1, \ldots, \lambda_n)$.
  The first $n$ coordinates are just  the eigenvalues of
$N$ whereas
 the remaining ones  can be constructed by quadratures
\cite{TMagri}.
The above form
will be referred to as the $canonical$ form of the previous BH
structure.

\subsection{Bi--Hamiltonian systems} \label{subsec: BHf}
 A vector field X is said to be bi-Hamiltonian  w.r.t.
 a pair of Poisson tensors $(P_0, P_1)$  if there exist
two smooth functions $h_0$ and $h_1$ such that

\begin{equation}  \label{eq:BHX}
 X=P_0\, dh_1=P_1\, dh_0 \ ,
\end{equation}
$d$ denoting the exterior derivative.
\par
If $N$ exists and it is maximal then a BH vector field is
completely integrable, a set of independent involutive integrals
being just  the eigenvalues $(\lambda_1, \ldots, \lambda_n)$
\cite{MM}. Conversely,   a  strong condition has to be satisfied
by a completely integrable system  in order to admit a BH
formulation  in a neighborhood of a Liouville  torus, at least if
one searches for a second Poisson tensor compatible  with  $P_0$
\cite{B,F}. Hereafter, this  kind of structures will
be referred to  as {\em standard}  bi--Hamiltonian
structures.

Nevertheless,  the property of Liouville integrability can be
related with  geometric formulations which are actually different
from the  BH one w.r.t. a standard BH structure. This can be done  in (at least)
three distinct ways.

\begin{itemize}
\item[i)]
Searching for a BH structure $(Q_0, Q_1)$  not including the 
Poisson tensor $P_0$ \cite{Mm1}; hereafter such  structures will
be referred to as  {\it alternative} BH structures.
\item[ii)]
Admitting a degenerate BH formulation;   this is the
case, for instance, of  the rigid body with a fixed point
\cite{Ratiu,  MP, Ian}  and of the
stationary flows of the KdV hierarchy \cite{AFW,GT1}.
\item[iii)]
Searching for a QBH formulation of the
vector field $X$ w.r.t. a standard BH structure 
\cite{francesi2,MT}.
\end{itemize}

\subsection{Quasi--bi--Hamiltonian systems} \label{subsec:
QBHf}
In connection with the third approach, we recall that the
vector field X is said to be
quasi-bi-Hamiltonian  w.r.t. a pair of Poisson tensors $(P_0, P_1)$ \cite{francesi2}
if there exist three smooth
functions $H$,   $K$,  $\rho$ such that

\begin{equation}  \label{eq:QBHX}
X=P_0\, dH=\frac{1}{\rho} P_1\, dK  \ .
\end{equation}
One could
say that the given dynamical system is Hamiltonian also
w.r.t. the Poisson tensor $P_1$, provided that
a nontrivial change in time: $dt\rightarrow d\tau=(1/\rho)
~dt$ is introduced.
In particular, a QBH vector field $X$ is said to be
Pfaffian if $N$ exists  and
$\rho=\prod_{i=1}^n\lambda_i $, i.~e., it is just the
product of the  eigenvalues of $N$.
The relevance of QBH vector fields is based upon the
following two  results.
\begin{itemize}
\item
 Any completely integrable system with two degrees of
freedom
admits a QBH formulation in a neighborhood of a Liouville
torus \cite{francesi2}.
\item
Any Pfaffian QBH vector field with $n$ degrees of freedom  is
separable (in the sense of Hamilton--Jacobi) in the
Darboux--Nijenhuis coordinates \cite{MT}. Indeed, the general
solution of Eq. (\ref{eq:QBHX}),  written  in these coordinates,
is

\begin{equation} \label{eq:HKnQBHS}
H=\sum_{i=1}^n\frac{f_i(\lambda_i;\mu_i)}{\prod_{\stackrel{\scriptstyle j=1}
{j\neq i}}^n(\lambda_i-\lambda_j)} \ , \quad
K=\sum_{i=1}^n \frac{\rho_i\,  f_i(\lambda_i;\mu_i)
}{\prod_{\stackrel{\scriptstyle j=1}
{j\neq i}}^n(\lambda_i-\lambda_j)}
 \qquad\quad 
 (\rho_i:=\prod_{\stackrel{\scriptstyle j=1}{j\neq i}}^n\lambda_j) \  ,
\end{equation}
 where  each function$f_i$ is an arbitrary smooth function,
depending at most on one pair of variables $(\lambda_i;\mu_i)$.
 (Obviously enough, if $n=1$ it is:
 ${\prod_{\stackrel{\scriptstyle j=1}{j\neq
i}}^n(\lambda_i-\lambda_j)}:=1$, $\rho_i=1$, $P_1=\lambda P_0$,
$H=K=f(\lambda, \mu)$). A remarkable feature of $H$ and $K$ is
that they are separable as they verify the Levi--Civita condition
\cite{LC}, so that the corresponding Hamilton equations are
integrable by quadratures. We stress the fact that, owing to the
arbitrariness of  $f_i$, the functions $H$ and $K$ (\ref
{eq:HKnQBHS}) provide a class of separable functions generally
different from the  known St\"{a}ckel class, quadratic in the
momenta (e.g., see  \cite[p. 101]{Per}).
\end{itemize}
 The  results of \cite{MT} have
been completed in \cite{Bl1}, where it has
been shown that a QBH vector field $X$ admits $n$ integrals
of motion in involution
$F_k \, (k=1,\ldots,n)$

\begin{equation} \label{eq:BlF}
 F_k= \sum_{i=1}^n \frac{\partial c_k}{\partial\lambda_i}
\frac{f_i(\lambda_i;\mu_i)}{\prod_{\stackrel{\scriptstyle j=1}
{j\neq i}}^n(\lambda_i-\lambda_j)}\ ,
\end{equation}
 $c_1,\ldots,c_n$ being the   coefficients of the minimal
polynomial of  $N$
\begin{equation} \label{eq:Nmp}
\lambda^n +\sum_{i=1}^{n}c_{i}\lambda^{n-i}=\prod_{i=1}^n
(\lambda-\lambda_i) \ ;
\end{equation} in particular,  $F_1=-H, F_n=(-1)^{n}K$.
Furthermore, each function
$F_k$ turns out to be  separable.

\vskip0.3truecm
{\bf Remark.}
Let $H$, $K$, $\rho$ be of the form (\ref{eq:HKnQBHS}),
i.e., the general
solutions of the QBH Eq.~(\ref{eq:QBHX}) in the Pfaffian
case. If $\varphi \ , \psi:
R\rightarrow R$  are $C^1$ functions with nonvanishing
derivatives $\varphi '$,  $\psi '$, let us define on the
phase space the functions
$\Phi$, $\Psi$ given by

\begin{equation}  \label{eq:HKrab}
\Phi (\mbox{\boldmath $\lambda$};\mbox{\boldmath  $\mu$}):=
\varphi (H (\mbox{\boldmath $\lambda$};\mbox{\boldmath
$\mu$})) \ , \qquad
\Psi (\mbox{\boldmath $\lambda$};\mbox{\boldmath  $\mu$}):=
\psi (K (\mbox{\boldmath $\lambda$};\mbox{\boldmath
$\mu$})) \ .
\end{equation}
Then  $\Phi$, $\Psi$ are solutions of the QBH Eq. (\ref{eq:QBHX})
with a function $\tilde{\rho}$ given by

\begin{equation}  \label{rhorab}
\tilde{\rho}=\rho \frac{\psi'(K)}{\varphi'(H)}  \ .
\end{equation}
Hence, for each QBH system of Pfaffian type there is a
class of QBH systems of non
Pfaffian type, with $\tilde{\rho}$ of the form
(\ref{rhorab}). Moreover, it is
straightforward to show that $\Phi$ and $\Psi$ are
separable, since they satisfy the
Levi--Civita condition, provided that  this condition be
fulfilled by $H$ and $K$ (as in
the Pfaffian case), whatever the particular form of the
functions $\varphi$ and
$\psi$.
\par
Viceversa,
let us search for the general solution of Eq.
(\ref{eq:QBHX}) w.r.t. $\tilde{H}$,
$\tilde{K}$ and
$\tilde{\rho}=\rho
\, f'(\tilde{H})/g'(\tilde{K})$ \  $(\rho=\prod_{i=1}^n
\lambda_i)$, with $f$ and $g$
arbitrarily chosen functions; it is easy to prove that
$\tilde{H}$ and  $\tilde{K}$ are
given by

\begin{equation} \label{eq: HtKt}
\tilde{H}=f^{-1}(H) \ , \qquad  \tilde{K}=g^{-1}(K)
\end{equation}
with $H$ and $K$ of the form (\ref{eq:HKnQBHS}).
\par
This kind of generalization of the Pfaffian case has been
recently considered in
\cite{Rab} only for the case $n=2$. As it is evident, the
above results hold for QBH
dynamical systems with an arbitrary  number of degrees of
freedom.
\rule{5pt}{5pt}

\subsection{ The ``origin" of some QBH vector fields}
\label{subsec: oQBH}
Let us describe a possible situation in which some interesting QBH
vector fields arise. Let  $(M, {\cal P}_0, {\cal P}_1)$ be a BH
manifold   and $X$ the vector field of a given dynamical system on
$M$, Hamiltonian w.r.t. ${\cal P}_0$. If neither ${\cal P}_0$ nor
${\cal P}_1$ is invertible, a possible way to analyse the
integrability of  $X$ is to eliminate the
Casimir functions of one Poisson tensor, say ${\cal
P}_0$,  by fixing their values; of course, both ${\cal P}_0$ and $X$ can  be
restricted to a symplectic leaf $S_0$ of ${\cal P}_0$, so that $X$ is still a
Hamiltonian vector field on $S_0$. However, if
 ${\cal P}_1$  cannot be restricted to $S_0$, the BH
formulation
(\ref{eq:BHX}) is lost on $S_0$, even if
$X$ is BH on $M$. As a matter of fact,  in a few cases the
following situation occurs:

\begin{itemize}
\item[i)]
 there is a fibration $\pi:M\rightarrow  M'=M/\pi$ such
that both
${\cal P}_0$  and
${\cal P}_1$ are projectable along $\pi$; since $\pi$ turns
out to be transversal to
$S_0$ and $S_1$ (a symplectic leaf of ${\cal P}_1$), the
quotient space
$M'$ and the  symplectic manifold $S_0$ are diffeomorphic
and   $S_0$ itself is a BH  manifold, with
$P_0$ and $P_1$ invertible, $P_0$ and $P_1$ denoting the
reduced tensors of
${\cal P}_0$ and ${\cal P}_1$ respectively. If this is the
case, and if the eigenvalues
of the Nijenhuis tensor
$N=P_1P_0^{-1}$ are  independent, then one can introduce
the Darboux-Nijenhuis
coordinates on $S_0$.
\item[ ii)]
 There is a function $\rho$ such that the restricted field
$X$ admits the  QBH
formulation (\ref{eq:QBHX}), with $H$ and $K$ given by the
restriction to $S_0$ of  integrals of motion of
$X$. The interest of this result is  that if  $X$ is
Pfaffian, then in the
Darboux-Nijenhuis  chart it is separable \cite{MT}.
\end{itemize}
We remark that the situation described in i) and ii) is, so
to say, ``experimental", a
sound theoretical foundation of these results being
lacking. The peculiarity of the
above reduction is given by the fact that  two different
geometric processes are
used simultaneously:  the restriction for the
  vector field and the  projection for the BH structure.
Due to this
fact,  one maintains the BH structure but loses the BH
formulation for the  vector field, recovering in
some cases the QBH formulation.  This happens, for
instance, for:

\begin{itemize}
\item
the integrable H\'enon--Heiles system and its
multidimensional generalizations
obtained by reduction from the {\it stationary} flows of
the KdV hierarchy
\cite{GT1, GT2};

\item
a class of permutationally symmetric potentials recovered from the
{\it restricted} flows of the coupled KdV systems, the most
representative member being the Garnier system \cite{Bl1}.

\end{itemize}
Both classes of dynamical systems live on a BH manifold $M$
of maximal rank
($dim M=2 n+1$) and their
 QBH formulation is obtained by the reduction to a
symplectic $2n$ dimensional manifold according to  the above
scheme i) , ii). The study of examples with BH structures of
non--maximal rank such  as the stationary flows of the Boussinesq
hierarchy will appear elsewhere \cite{FMT}. In this regard, we
recall that the  geometry of $(2 n+1)$-dimensional BH manifolds of
maximal rank   has been completely described by Gelfand and
Zakharevich \cite{GZ}, whereas there is not  a similar analysis for
non--maximal  BH manifolds.
\par
For the sake of clarity, we recall that we are considering from
the very beginning  a well specified vector field $X$ which is
associated to the dynamical system under investigation. So, the
situation described above does not contradict  the following known
results: if one has been given a BH structure  with $P_0$
invertible, one can construct a BH vector field whose Hamiltonian
functions are  the  traces (or the eigenvalues) of the Nijenhuis
tensor $N$ \cite{MM};   moreover,  a QBH vector field can always
be constructed for any maximal Nijenhuis tensor 
\cite{FMT}. However, it turns out that these BH and
QBH  fields do not in general coincide with the vector field $X$
of the given dynamical system.

\section{Examples of dynamical systems with QBH
formulation} \label{sec:e}
 We present four examples of   integrable systems  which can be
given a QBH formulation of Pfaffian type w.r.t. a standard BH structure.
\par
   The first three systems  (see, e.g.,
   \cite[p.126--129]{Ar}), are
defined on the cotangent bundles of  Riemannian manifolds;
therefore they are of  St\"{a}ckel type and the Nijenhuis tensor
of their QBH formulation can be constructed by lifting, in a
suitable way, a {\em conformal Killing tensor} \cite{Ben} from
the
 configuration space  to the corresponding cotangent bundle.
 (Such a construction will be the subject of a further
publication).
 The dynamical system considered in the last example is naturally
Pfaffian QBH,   the physical coordinates being just 
Darboux-Nijenhuis coordinates w.r.t. a standard BH structure written in the canonical form
(\ref{eq:PND}).

\subsection{The Kepler problem with a homogeneous force
(Lagrange 1766)}
\label{subsec:Ks}
Let us consider the classical problem of a particle in the
plane under the influence of
the Kepler potential  and of a homogeneous field force.
The Hamiltonian function is

\begin{equation} \label{eq:HKs}
H=\frac{1}{2}(p_1^2+p_2^2)-\frac{a}{\sqrt{q_1^2+q_2^2}}-b
q_2 \ ,
\end{equation}
where $(q_1,q_2;p_1,p_2)$ are the cartesian coordinates of
the particle and the
conjugate momenta, respectively;  $a$ and $b$ are real
constants. The   vector
field is $X=P_0\, dH$,  $P_0$ being the canonical Poisson
tensor. There is  a second independent integral
of motion

\begin{equation} \label{eq:KKs}
K=\frac{ p_1}{2}(q_2 p_1-q_1 p_2)-\frac{a
q_2}{2\sqrt{q_1^2+q_2^2}}+
\frac{bq_1^2}{4} \ ,
\end{equation}
which allows us to give $X$ a QBH formulation. Indeed, we
can write $X=\frac{1}{\rho}   P_1\, dK$ with
$\rho=-\frac{q_1^2}{4}$ and

\begin{equation}   \label{eq:P1Ks}
P_1=\frac{1}{2}\left[
\begin{array} {cccc}
0&0&0&q_1 \\
0&0&q_1&2q_2 \\
0&-q_1&0&-p1 \\
-q_1&-2q_2&p_1&0
\end{array}
\right]
\end{equation}
Since  the minimal polynomial of the Nijenhuis tensor
$N=P_1P_0^{-1}$ is

\begin{equation}  \label{mpNKs}
m(\lambda)= \lambda^2 -q_2 \lambda -\frac{q_1^2 }{4} \ ,
\end{equation}
one easily checks that $X$ is a Pfaffian QBH vector field.
The Darboux--Nijenhuis
coordinates can be constructed following \cite{TMagri};
they are

\begin{equation} \label{DNcKs}
\lambda_{1,2}=\frac{1}{2}(q_2\mp \sqrt{q_1^2+q_2^2}) \qquad
\mu_{1,2}=p_2-\frac{p_1}{q_1}(q_2 \pm \sqrt{q_1^2+q_2^2}) \
.
\end{equation}
One can easily verify that $(\lambda_1,\lambda_2)$ are just
parabolic
coordinates in the plane with focus at the point
$q_1=q_2=0$ and axis the $q_2$-axis.
In the above coordinates the two integrals of motion read

\begin{eqnarray}
H &=&
\frac{\lambda_1\mu_1^2-2b\lambda_1^2+a}{2(\lambda_1-\lambda
_2)}+
\frac{\lambda_2\mu_2^2-2b\lambda_2^2+a}{2(\lambda_2-\lambda
_1)} \\
K &=&\lambda_2
\frac{\lambda_1\mu_1^2-2b\lambda_1^2+a}{2(\lambda_1-\lambda
_2)}+
 \lambda_1
\frac{\lambda_2\mu_2^2-2b\lambda_2^2+a}{2(\lambda_2-\lambda
_1)}
\ .
\end{eqnarray}
On account of the general result proved in \cite{MT}, one
recovers that $H$ and $K$ are separable.

\subsection{The Euler problem with the two fixed centers
and an elastic force (Euler
1760--Lagrange 1766)}
\label{subsec:Es}

The Hamiltonian function for this problem can be written as
follows

\begin{equation} \label{eq:HEs}
H=\frac{1}{2}(p_1^2+p_2^2)+\frac{b_1}{\sqrt{q_1^2+(q_2+c)^2
}}+
     \frac{b_2}{\sqrt{q_1^2+(q_2-c)^2}} +\frac{
k}{2}(q_1^2+q_2^2)
\ ,
\end{equation}
where: $(q_1,q_2; p_1,p_2)$ are as in the previous example; the
two centers are at the points $F_1=(0,-c)$ and $F_2=(0,c)$;
 $b_1$, $b_2$, $k$ are real constants.   The corresponding
vector field is $X=P_0\, dH$. As above,
also this system does not admit a BH formulation w.r.t. a
second Poisson tensor $P_1$, but it can be given a
QBH formulation. Indeed, one can write Eq.
(\ref{eq:QBHX})  with $K$ given by the
integral of motion ($a_1 \in  R$, $a_2=a_1+c^2$)

\begin{equation} \label{eq:KEs}
K=\frac{a_2}{2}p_1^2+\frac{a_1}{2}p_2^2-\frac{1}{2}
(q_1p_2-q_2
p_1)^2+
\frac{b_1(a_2+c q_2)}{\sqrt{q_1^2+(q_2+c)^2}}+
\frac{b_2(a_2-c q_2)}{\sqrt{q_1^2+(q_2-c)^2}}
+\frac{ k}{2}(a_2 q_1^2+a_1q_2^2)
\ ,
\end{equation}
$\rho=(a_1 a_2-a_2 q_1^2-a_1q_2^2) $   and the
following Poisson tensor $P_1$

\begin{equation}   \label{eq:P1Es}
P_1=\left[
\begin{array} {cccc}
0&0&a_1-q_1^2&-q_1q_2 \\
0&0&-q_1q_2&a_2-q_2^2 \\
-(a_1-q_1^2)&q_1q_2&0&-q_1p_2+q_2p_1 \\
q_1q_2&-(a_2-q_2^2 )&q_1p_2-q_2p_1&0
\end{array}
\right]  \ .
\end{equation}
The minimal polynomial of the Nijenhuis tensor
$N=P_1P_0^{-1}$ is

\begin{equation}  \label{mpNEs}
m(\lambda)= \lambda^2 +(q_1^2+q_2^2-a_1-a_2)\lambda+
a_1 a_2-a_2 q_1^2-a_1q_2^2 \ ,
\end{equation}
so one easily recognizes the Pfaffian property of $X$.  The
Darboux--Nijenhuis coordinates are given
by

\begin{eqnarray}\label{DNcEs}
\lambda_{1,2}&=&\frac{(a_1+a_2-q_1^2-q_2^2)}{2}\mp
\frac{\sqrt{(q_1^2+q_2^2-a_1-a_2)^2-4(a_1
a_2-a_2q_1^2-a_1q_2^2)}}{2}  \\
\mu_{1,2}&=&-\frac{p_1q_2+p_2q_1}{4q_1q_2}+
\frac{(p_1q_2-p_2q_1)(q_1^2+q_2^2)}{4(a_2-a_1)q_1q_2}
\nonumber \\
& \mp& \frac{(p_1q_2-p_2q_1)}{4(a_2-a_1)q_1q_2}
\sqrt{(a_2-a_1)(a_2-a_1+2q_1^2-2q_2^2)+(q_1^2+q_2^2)^2} \ .
\end{eqnarray}
One can verify that $(\lambda_1,\lambda_2)$ are just the
Jacobi elliptic
coordinates in the plane (already known to Euler) with foci
at the points
$F_1$, $F_2$ and axis
the $q_2$-axis.  In the above coordinates the two integrals
of motion read

\begin{eqnarray}
H&=&\frac{-2(a_1-\lambda_1)(a_2-\lambda_1)\mu_1^2-(b_1+b_2)
\sqrt{a_2-\lambda_1}-
\frac{k}{2}\lambda_1^2} {\lambda_1-\lambda_2}  \nonumber \\
&+&
\frac{-2(a_1-\lambda_2)(a_2-\lambda_2)\mu_2^2+(b_2-b_1)\
sqrt{a_2-\lambda_2}-
\frac{k}{2}\lambda_2^2}
{\lambda_2-\lambda_1}
  \\
K&=&\lambda_2\frac{-2(a_1-\lambda_1)(a_2-\lambda_1)\mu_1^2-
(b_1+b_2)\sqrt{a_2-\lambda_1}-
\frac{k}{2}\lambda_1^2} {\lambda_1-\lambda_2}  \nonumber \\
&+&
\lambda_1\frac{-2(a_1-\lambda_2)(a_2-\lambda_2)\mu_2^2+(b_2
-b_1)\sqrt{a_2-\lambda_2}-
\frac{k}{2}\lambda_2^2}
{\lambda_2-\lambda_1}
\end{eqnarray}
Also in this case they are separable, since they have the
general form (\ref{eq:HKnQBHS}).

\subsection{The motion on the ellipsoid in an elastic
potential (Jacobi 1843)}
\label{subsec:gf}
Let us consider   a harmonic oscillator  on the
$n$-dimensional ellipsoid

\begin{equation} \label{Je}
\sum_{i=0}^{n}{\frac{x_i^2}{a_i}}=1 \ ,
\end{equation}
where $x_i \, (i=0,1,\cdots ,n)$ are Cartesian coordinates in
$R^{n+1}$ and $a_0< a_1< a_2< ~\cdots~<~ a_n$ are positive real
constants. If $y_i$ are  the conjugate momenta, the Hamiltonian
function is $H=\frac{1}{2}\sum_{i=0}^{n}(y_i^2+x_i^2)$. Let us
consider  the canonical transformation $(x_i,y_i)\mapsto
(\lambda_i,\mu_i)$ associated to the point transformation
$(x_i)\mapsto (\lambda_i)$, where $\lambda_i$ are the generalized
elliptic coordinates
 in $R^{n+1}$ \cite{Ar}  defined as  the $ (n+1)$ roots of
the equation

\begin{equation}
\sum_{i=0}^{n}{\frac{x_i^2}{a_i-\lambda}}=1 \ .
\end{equation}
 The ellipsoid (\ref{Je}) is given by the submanifold
$\lambda_0=0$,  and   the Hamiltonian function of the
harmonic oscillator
restricted to the cotangent bundle
$\lambda_0=0,\mu_0=0$   reads  (up to an
inessential constant term)

\begin{equation} \label{eq:Helli}
H=\frac{1}{2}\sum_{i=1}^{n}
\frac{4\prod_{j=0}^{n}(a_j-\lambda_i)\lambda_i^{-1}\mu_i^2+
k\lambda_i^{n}}
{\prod_{\stackrel{\scriptstyle j=1}{j\neq i}}^n
(\lambda_i-\lambda_j)}
\ ,
\end{equation}
where ${\prod_{\stackrel{\scriptstyle j=1}{j\neq
i}}^n(\lambda_i-\lambda_j)}:=1$ for $n=1$, (recall 
that $\sum_{i=1}^n \lambda_i^n/
{\prod_{\stackrel{\scriptstyle j=1}
{j\neq i}}^n}(\lambda_i-\lambda_j)$ can be written as $\sum_{i=1}^n\lambda_i$).
Since $H$ has the form (\ref{eq:HKnQBHS}), the general result
proved in \cite{MT} allows us to infer immediately that

\begin{itemize}
\item[i)]
the elliptic coordinates and the corresponding momenta
$\mu_i$ are  Darboux--Nijenhuis
coordinates for a  standard BH structure $(P_0,P_1)$;
\item[ii)]
in these coordinates, $P_1$ takes the canonical form
(\ref{eq:PND});
\item[iii)]
the vector field $X=P_0\, dH$ is a Pfaffian QBH vector
field. The function $K$ takes
the form
\begin{equation}
K=\frac{1}{2}\sum_{i=1}^{n}
\rho_i
\frac{4\prod_{j=0}^{n}(a_j-\lambda_i)\lambda_i^{-1}\mu_i^2+
k\lambda_i^{n}}
{\prod_{\stackrel{\scriptstyle j=1}{j\neq
i}}^n(\lambda_i-\lambda_j)} \ ;
\end{equation}
\item[iv)]
a complete set of rational integrals of motion in
involution is given by   (\ref{eq:BlF}).
\end{itemize}

\subsection{A  solvable $n$--body problem}
This system belongs to a large class of integrable
$n$--body problems in the plane, recently introduced by F.
Calogero \cite{Cal}.  Let
$M=C^{2n}$ (with  coordinates
$ \mbox{\boldmath $\lambda$}:=(\lambda_1,\ldots,\lambda_n)$
and the conjugate
momenta $\mbox{\boldmath  $\mu$}:=(\mu_1,\ldots,\mu_n)$)
be  the phase space of
the dynamical system with Hamiltonian

\begin{equation} \label{eq:CH}
H=\sum_{i=1}^n \frac{g_i(\lambda_i) e^{a\mu_i}+b
\lambda_i^n}
{\prod_{\stackrel{\scriptstyle j=1}{j\neq
i}}^n(\lambda_i-\lambda_j)}  \ ,
\end{equation}
where $g_i$ are arbitrary smooth functions, each one
depending only on the corresponding
coordinate
$\lambda_i$, and
$a$, $b$ are  arbitrary  constants.
 The related   Newton equations of motion take the form

\begin{equation}  \label{eq:Ne}
\ddot \lambda_k=2 \sum_{i\neq k}
\frac{\dot \lambda_i \dot\lambda_k}{\lambda_k-\lambda_i}
-b\dot \lambda_k \ .
\end{equation}
In the  case $b=0$,  a  QBH formulation and an alternative
BH formulation
(according to the item i) of Subsec. \ref{subsec: BHf}) has
been discussed in
\cite{MTcal}, while  in  the case $b=\sqrt{-1}\,\omega$,
($\omega \in R$)
the motion has been proved to be completely periodic   in
\cite{Cal}.

\par
Just as in the  previous example of this  section,   by
comparing (\ref{eq:CH}) with
(\ref{eq:HKnQBHS}) one immediately concludes that:

\begin{itemize}
\item[i)]
the  coordinates $\lambda_i$   and the corresponding momenta
$\mu_i$ are  Darboux--Nijenhuis coordinates for a  standard  BH
structure $(P_0,P_1)$;

\item[ii)]
in these coordinates, $P_1$ takes the canonical form
(\ref{eq:PND});
\item[iii)]
the vector field $X=P_0\,dH$ is a Pfaffian QBH vector
field. The function $K$ takes
the form

\begin{equation} \label{eq:CK}
K=\sum_{i=1}^n \rho_i \frac{ g_i(\lambda_i)e^{a\mu_i} +b
\lambda_i^{n}}
{\prod_{\stackrel{\scriptstyle j=1}{j\neq i}}^n(\lambda_i-\lambda_j)} \ .
\end{equation}

\item[iv)]
a complete set of  integrals of motion in involution is
given by   (\ref{eq:BlF}).
\end{itemize}

Furthermore, the
corresponding Hamilton--Jacobi equation
  is separable; a complete
integral is
$S(\mbox{\boldmath $\lambda$}; b_1,\ldots
,b_n)=-b_1t+W(\mbox{\boldmath
$\lambda$};b_1,\ldots ,b_n)$  with

\begin{equation} \label{eq:W}
W=\frac{1}{a} \sum_{i=1}^n    \int^{\lambda_i}
\log\left(\frac{1}{g_i(\xi)}\sum_{j=0}^n
b_j
\, \xi^{n-j}\right) d\xi \ ,
\end{equation}
with $b_0=b$ and $b_1=H$.
\par
We wish to stress that, unlike  the previous three examples,
this system is not of
St\"{a}ckel type, nevertheless it is separable as well.

\section{Concluding remarks}
The Kepler and the Euler problems considered in Subsections
\ref{subsec:Ks} and \ref{subsec:Es} correspond to Hamiltonian
systems with two degrees of freedom. We recall that for such kind
of systems there is a general result stating that a QBH
formulation always exists \cite{francesi2}: however, it is
essentially different from those presented above. Hence, the two
systems are explicit examples of the non uniqueness of the QBH
formulation.
\par
Indeed, in \cite{francesi2} one assumes to have a vector field
$X$, Hamiltonian w.r.t. an invertible Poisson tensor and
Liouville--integrable; passing to the action--angle variables, one
can conclude that, for $n=2$,  Eq. (\ref{eq:QBHX}) holds for a
suitable Poisson tensor $P_1'$, a function $K'$ and an integrating
factor  $\rho'$. In particular, $\rho'$ depends only on the action
variables, so it is a constant of motion for $X$; this general
property allows us to infer, by simple inspection, that our QBH
formulations are different, since in both cases the functions
$\rho$ depend only on the configuration variables and therefore
they are not  integrals of motion for the corresponding dynamical
systems.
\par
Finally, let us make a few comments about  the relevance of the
QBH formulation  (we thank an anonymous referee for arising this
question).
\par
 Let us consider two different situations (as well as those presented in
Subsect. \ref{subsec: oQBH} and Sect.  \ref{sec:e}, respectively),
reflecting  two ways in which the QBH systems arise;  these
situations can be classified according to the fact that the phase
space $M$ of the given dynamical system $X$ is:

\begin{itemize}
\item[i)]
 a BH  manifold $(M, {\cal P}_0,{\cal P}_1)$, with both
${\cal P}_0$ and ${\cal P}_1$
non invertible;   
\item[ii)]
  a symplectic manifold $(M, P_0=\omega_0^{-1})$,
$\omega_0$  being a symplectic tensor.
\end{itemize}

In the first case,
 if one wants  to solve the equations of motion through
 a complete integral of the
Hamilton--Jacobi   equation,  then one has  to pass to a
symplectic manifold (as it happens for the stationary \cite{DKN}
and the restricted flows of the KdV hierarchy, whose phase space is
odd--dimensional). As a matter of fact, the symplectic manifold
 is the proper geometrical setting where the
Hamilton--Jacobi method must be set up \cite{MHJ} (after having
been usually considered on   cotangent bundles). To the best of
our knowledge, a further generalization to Poisson (not
symplectic) manifolds is still lacking. A possible way to achieve
this goal    is to perform a ``reduction procedure"  to a
symplectic leaf $S_0$ of one of the Poisson tensor, getting (if
any) a QBH formulation for  $X$  and a Nijenhuis tensor
$N:=P_1P_0^{-1}$ (see  Subsect. \ref{subsec: oQBH} for details and
notations). It is just such a tensor,  living  on $S_0$ and not
existing on    $M$,   that allows us to define in an {\em
intrinsic}  way, by means of  its spectral data,
 the Darboux--Nijenhuis coordinates.    We recall that such
coordinates
are separation
 variables for the Hamilton--Jacobi equation corresponding
to  any Pfaffian QBH
system.
\par
In the second case,  $M$ is a symplectic manifold. If $X$ is
Liouville--integrable, it always admits (infinitely many) {\em
alternative} BH formulation,  not  including $P_0$  (as recalled
in item i) of Subsect. \ref{subsec: BHf}). However, if one wants
still to exploit $P_0$ for getting a standard BH structure, in many cases  a QBH formulation
for $X$ can be constructed in a quite natural way, as it has been shown
  in the examples discussed above.
 Of course,  in this case one could also try to embed the given dynamical
system $X$ in   a larger phase space and to lift the QBH
formulation  in order to get a BH formulation for
(the lifting of) $X$, somehow reverting  the above
mentioned reduction procedure. However,   this lifting is neither
natural nor  unique, and would oblige  to work anew in  a Poisson
(not symplectic) manifold.
\vspace{1truecm}
\par\noindent
{\bf Acknowledgments.}
One of us (G.T.)   thanks   the organizers for the kind
hospitality
in Torun during the XXX Symposium on Mathematical
Physics,   where a shorter
version of this report was presented.  Moreover,
enlightening discussions with
M. B{\l}aszak,  F.~Calogero,   A. P. Fordy, J.-P.
Francoise  and  A. P. Veselov  are
gratefully acknowledged.


\begin{thebibliography}{A}



\bibitem[1]{francesi2} R. Brouzet, R. Caboz, J. Rabenivo,
V. Ravoson,  J. Phys. A {\bf 29} (1996), 2069--2076.



\bibitem[2]{MT}
 C. Morosi, G. Tondo,
 J. Phys. A {\bf 30} (1997), 2799-2806.

\bibitem[3]{francesi1}
R. Caboz, V. Ravoson, L. Gavrilov, J. Phys. A {\bf 24} (1991),
L523--L525.



\bibitem[4]{Bl1}  M. B{\l}aszak,
  J. Math. Phys. {\bf 39} (1998), 3213--3235.

\bibitem[5]{Bl2} M. B{\l}aszak,
 Phys. Lett. A. {\bf 243} (1998), 25--32.

 \bibitem[6]{MTcal}
 C. Morosi, G. Tondo, Phys. Lett. A {\bf 247} (1998), 59--64.


\bibitem[7]{Cal} F. Calogero,  J. Math. Phys. {\bf 38} (1997),
5711--5719.

\bibitem[8]{Magri1}
 F. Magri,  J. Math. Phys.{\bf 19} (1978), 1156--1162.


\bibitem[9] {MagriL} F. Magri, {\it A Geometrical approach
to the nonlinear solvable equations,}in
 Lecture Notes in Physics {\bf  120}, (M. Boiti,
F.~Pempinelli, G.~Soliani eds.),
 Springer--Verlag, Berlin 1980, pp. 233--263.


\bibitem[10]{FoFu} A. Fokas, B. Fuchssteiner,
  Physica D{\bf 4} (1981), 47--66.




\bibitem[11]{TMagri} F. Magri, T. Marsico, {\it Some
developments of the concepts of
Poisson manifolds in the sense of A. Lichnerowicz,} in
Gravitation, Electromagnetism
and Geometric Structures, (G. Ferrarese, ed.), Pitagora
editrice, Bologna  1996, pp.
207--222.



\bibitem[12]{MM}  F. Magri, C. Morosi, {\it A Geometric
Characterization of Integrable
Hamiltonian Systems through the Theory of
Poisson--Nijenhuis Manifolds,}
Quaderno {\bf S/19} (1984), Universit\`a di Milano.

\bibitem[13]{B}
 R. Brouzet, J. Math. Phys. {\bf  34} (1993), 1309--1313.

\bibitem[14]{F} R. L. Fernandes,
    J. Dynam. Differential
Equations {\bf 6} (1994), 53--69.


\bibitem[15]{Mm1} S. De Filippo, G. Vilasi, G. Marmo, M.
Salerno, Nuovo Cimento B {\bf 83} (1984), 97--112.




\bibitem[16]{Ratiu}  T. Ratiu,  Amer. Jour. Math. {\bf
104} (1982), 409--448.


\bibitem[17]{MP} C. Morosi, L. Pizzocchero,
Lett. Math. Phys. {\bf 37}
(1996), 117--135.

\bibitem[18]{Ian} I. D. Marshall,  Comm. Math. Phys. {\bf 191}
(1998), 723--734.

\bibitem[19]{AFW} M. Antonowicz, A. P. Fordy, S.
Wojciechowski,  Phys. Lett. A {\bf 124} (1987), 455-462.

\bibitem[20]{GT1} G. Tondo,   J. Phys. A {\bf 28} (1995),
5097--5115.



\bibitem[21]{LC} T. Levi-Civita,  Math. Ann.
{\bf 59} (1904),
383--397.


\bibitem [22] {Per} A.M. Perelomov, {\it Integrable systems of classical
Mechanics and Lie algebras,} Birkh\"{a}user, Berlin 1990.



\bibitem[23]{Rab} J. Rabenivo,
J. Phys. A  {\bf 31} (1998), 7113--7120.



\bibitem[24]{GT2}
 G. Tondo,   {\it On the integrability of H\'enon--Heiles
type systems,} in
  Non Linear Physics, Theory and Experiment,
 (E. Alfinito et al. eds.), World Scientific,  Singapore
1996,
pp. 313--320.


\bibitem[25]{FMT}
 G. Falqui,   F. Magri, G. Tondo, {\it Reduction of
bihamiltonian systems and
separation of variables: an example from the Boussinesq
hierarchy,} in
Proceedings of  NEEDS in LEEDS,   submitted to Theor. Math. Phys.
(1998).

\bibitem[26]{GZ} I.M. Gelfand, I. Zakharevich, {\it On the
local geometry of a
bi--Hamiltonian structure,} in The Gelfand Mathematical
Seminars 1990--92, (L.
Corwin et al., eds.), Birk\"{a}user, Boston 1993, pp. 51--112.


\bibitem[27]{Ar} V.I.  Arnold, V.V. Kozlov, A.I. Neishtadt,
{\it Mathematical Aspects
of Classical and Celestial Mechanics,} in Dynamical Systems
III, EMS vol.~3,
Springer--Verlag, Berlin  1988.


\bibitem[28]{Ben} S. Benenti,
J. Math. Phys. {\bf 12} (1997), 6578--6602.


\bibitem[29]{DKN}
B. A. Dubrovin, I. M. Krichever, S. P. Novikov, {\em
Integrable Systems I,} in Dynamical Systems IV, EMS vol.~ 4,
 Springer--Verlag,  Berlin 1990, pp. 173--280.


\bibitem[30]{MHJ} G. Marmo,
Riv. Nuovo Cimento
{\bf 13} (1990), 1--74.




\end{thebibliography}
\end{document}